\documentclass[12pt,letterpaper,fleq,onecolumn]{article}
\usepackage[dvips]{graphics,color,epsfig} 
\usepackage{bm} 

\begin{document}
\baselineskip 18 true pt

\def\aa{{\alpha_1}}
\def\ab{{\alpha_2}}
\def\H{{H$\ddot{\rm o}$lder\ }}
\def\T{{\it T}_\psi}
\def\V{{\it V}}
\def\A{{\bf A}}
\def\F{{\bf F}}

\def\pa{p_0}
\def\pb{p_1}
\def\ma{m_0}
\def\mb{m_1}

\def\Z{{\cal Z}}
\def\pdf{{\tt PDF}}
\def\F{{\bf F}}\def\w{{\omega}}

\title{Wavelet Transform Modulus Maxima Based \\
Fractal Correlation Analysis}

\author{D.C. Lin$^1$, A. Sharif\\
$^1$Department of Mechanical and Industrial Eng., Ryerson University,\\
Toronto, ON, Canada\\}

\maketitle

\begin{abstract}

The wavelet transform modulus maxima (WTMM) used in the singularity analysis
of one fractal function is extended to study the fractal correlation of two
multifractal functions. The technique is developed in the framework of joint
partition function analysis (JPFA) proposed by Meneveau et al. [1] and is
shown to be equally effective. In addition, we show that another leading
approach developed for the same purpose, namely, relative multifractal
analysis, can be considered as a special case of JPFA at a particular
parameter setting.

\end{abstract}

\section{Introdcution}

Fluctuations in many natural and artificial phenomena are found to exhibit
fractal characteristics. In applications, this has been characterized by
the so-called singularity spectrum of some numerical or experimental data.
To understand the dynamics underlying the fractal, it is not uncommon that
multiple data capturing different aspects of the phenomenon of interest are
used in the analysis. For example, velocity and temperature flucutations
are used to analyze the momentum and energy aspects of the multifractal
hydrodynamic turbulence [1], blood pressure and heart rate fluctuations to
analyze the cardiovascular aspect of the $1/f$-like power spectrum of the
heart rate variability in humans [2], packet size and arrival time to
analyze the congestion and connectivity aspects of the multifractal network
traffic [3], and so on. For fractal analysis on natural objects, multiple
data cross examination may provide the chance to examine the potential link
of fractal fluctuation in the data. In particular, one would suspect some
degree of {\it fractal correlation} in the data if the fractal generating
mechanisms associated with the data source are coupled together.

Essential to the notion of fractal correlation is the distinguishability
of singularity spectra. There are fundamental and practical issues related
to this subject. For example, consider a standard $N_p$-adic multinomial
process on an interval. It is a multiplicative cascade constructed by
repeatedly dividing the interval into equal $N_p$ segments and assigning
(probability) weights $p_i,i=1,\cdots,N_p$, from one generation to the
next. Continuing this procedure {\it ad infinitum} leads to a limiting
process with no density (almost surely) and intermittent spiking pattern.
Its singularity spectrum $f_\pi(\alpha)$ may be estimated by the Legendre
transform of $\tau_\pi(q)=-\log(\sum p_i^q)/\log(N_p)$:
$$f_\pi(\alpha)=q\alpha-\tau_\pi(q)$$
where $\alpha(q)=d\tau_\pi/d q$. Now, consider a different $N_m$-adic
cascade ($N_m\ne N_p$) generated by weights $m_i,i=1,\cdots,N_m$, and its
singularity spectrum $f_\mu(\alpha)$. If $f_\pi(\alpha)=f_\mu(\alpha)$,
one must have
$$\sum p_i^q=\left(\sum m_i^q\right)^{\log(N_p)/\log(N_m)}.$$
However, no $\{p_i\}$ and $\{m_i\}$ can be found to satisfy this equation
for all $q$. Thus, singularity spectra can in theory be distinguished, at
least for the important class of multinomial processes. For more in-depth
treatments and examples, see the excellent book by Pesin [4].

In practice, a different issue can arise. That is, two singularity spectra
may be close to each other within the limit of finite precision. Consider
again the cascades from above. Let $\tau_\pi(-\infty)=\tau_\mu(-\infty)$,
$\tau_\pi(+\infty)=\tau_\mu(+\infty)$ (so $\max(\{p_i\})=\max(\{m_i\})^{
\log(N_p)/\log(N_m)}$, $\min(\{p_i\})=\min(\{m_i\})^{\log(N_p)/\log(N_m)}$).
Then, $f_\pi(\alpha),f_\mu(\alpha)$ will agree at four important $q$
values: $-\infty,+\infty,0,1$. With the rest of $p_i$ and $m_i$ chosen
properly, they can be made almost indistinguishable (Fig.~1). This problem
was addressed by L${\acute e}$vy-L${\acute e}$hel and Vojak who developed
a much sharper mutual multifractal analysis to relate the singularity
spectra to the generation of multinomial processes [5]. Riedi and Scheuring
arrived at the similar relative multifractal analyses with further details
on the numerical implementation [6]; see also [7, 8]. For experimental
data, Meneveau et al. introduced a joint partition function analysis (JPFA)
based on the 1D version introduced by Hentschel and Procaccia [9] and
Halsey et al. [10]. With essentially the same procedure of estimating the
singularity spectrum, these authors characterized fractal correlation in
the small scale kinetic energy transfer, heat concentration and vorticity
of the turbulent flow [1]. They implied that multiple cascades of more
than one variables are responsible for the fractal fluctuation in fluid
turbulence; see also [11]. The JPFA has also been applied in diversed areas
such as precision agriculture [12], soil property [13], and re-emerged in
the general discussion of discrete scale invariance of the multinomial
process [14].

In these past studies, JPFA as well as most singularity analyses were
conducted on the assumption of one-dimensional multifractal measures and
solved using the classical box counting procedure. Such applications can
be limited in scope when, typically in the experimental study, the data is
a fractal function that is not additive. For functions created by the
integral of multinomial measure under $C^\infty$ perturbation [15], Bacry
et al. proved that the singularity analysis developed for fractal measures
is equally applicable by using the so-called wavelet transform modulus
maxima (WTMM) method. Later, the validity of WTMM was examined by Jaffard
for any function [16]. Regarding WTMM, it was proved that (i) it can yield
the upper bound estimate of the singularity spectrum for any function and
(ii) it is exact for the so-called self-similar fractal function, as long
as the so-called maxima lines are not too close to each others.

The purpose of this study is to introduce a joint WTMM method to carry out
the JPFA of fractal correlation. The WTMM-based approach can provide an
accessible and stable tool for estimating fractal fluctuation in multiple
experimental data. We also point out that JPFA is a more general formulation.
In particular, the existing relative multifractal analysis developed for
the similar purpose relates to JPFA at a particular parameter setting. While
the term correlation is normally linked to the second order statistics,
fractal correlation as estimated from the singularity spectra is a property
of moment of all orders. Indeed, the primary object of the analysis is the
Hausdorff dimension $f(\aa,\ab)$ of the support of observing \H exponents
$\aa$ and $\ab$. It will be shown that the set $f(\aa,\ab)$ describes a
two-dimensional surface whose level sets characterize the coupling of
fractal generating mechanisms as well as the relative multifractal spectrum
studied in the past.

Our results are organized into four sections. In the next section, the
background of WTMM is first summarized. The extension to the WTMM-based
JPFA and its connection to the relative multifractal spectrum are then
given. The test of the method using multinomial cascades are presented
in section 3. Concluding remarks are given in section 4.

\section{WTMM--Based Fractal Correlation Analysis}

\subsection{WTMM Singularity Analysis}

Singularity analysis is built on the notion of \H continuity of functions.
Recall that a function $x(t)$ is \H continuous of exponent $\alpha'$ if
there are $\alpha', \delta_0, C\in{\bf R}^+$, such that, for $\delta<\delta
_0$,
$$|x(t_0+\delta)-x(t_0)|\le C|\delta|^{\alpha'}.$$
In the neighborhood of $x(t_0)$, there exists a supremum $\alpha(t_0)$
that (1) is valid for all $\alpha'\le\alpha(t_0)$. The exponent $\alpha(
t_0)$ is the \H exponent of $x(t)$ at $t_0$. Formally [15, 16], one can
find an $n$th order polynomial $P_n(t)$ and $\alpha(t_0)\in[n,n+1)$ such
that
$$|x(t_0+\delta)-P_n(t_0)|\le C|\delta|^{\alpha(t_0)}.
\eqno(1)$$
It is evident that the \H exponent characterizes the differentiability of
the function and, thus, how the function can fluctuate. For example,
$\alpha=+\infty$ for $C^\infty$ functions, $\alpha\in(n,n+1)$ for functions
that are only $n$ times differentiable and $\alpha<1$ for functions that
are non-differentiable. The $\alpha<1$ case draws the most attention since
it means the function can fluctuate in large amplitude over short time
intervals and gives rise to the so-called intermittent pattern witnessed in
many physical systems. In this case, the \H exponent $\alpha(t)$ is also
known as the singularity exponent.

The natural tool to analyze the singularity property is by the wavelet
transform:
$$\T[x](t,a) = {1\over a}\int_{-\infty}^\infty\psi\left({t'-t\over a}\right)
x(t')dt'
\eqno(2)$$
where $\T[x](t,a)$ is the wavelet coefficient and $\psi(t)$ is the analyzing
wavelet. Muzy et al. showed that the exponent $\alpha(t)$ can be estimated
effectively using the supremum of $|\T[x]|$ along the so-called maxima line
formed by the local wavelet modulus maxima [18]. Denote the set of maxima
lines at scale $a$ by ${\cal L}(a)=\{l_1,l_2,\cdots,l_{N(a)}\}$. Bacry et al.
proved that [15]
$$Z(a;q)=\sum_{l_i\in{\cal L}(a)}C_i^q\sim a^{\tau(q)}
\eqno(3)$$
where $C_i=\sup_{(t,a)\in l_i}|\T[x](t,a)|$ is the supremum of the modulus
maxima along the maxima line $l_i$. For functions created by the integral
of multinomial measures under $C^\infty$ perturbations, it was shown that
the Legendre transform of $\tau(q)$ yields the Hausdorff dimension of
the support $\{t,\alpha(t)=\alpha\}$, $f(\alpha)$; 
$$\tau(q)=\min_\alpha(q\alpha - f(\alpha));
\eqno(4)$$
see also [16]. In the literature, $Z(a;q)$ is sometimes referred to as the
partition function due to its analogy to the energy partition function in
statistical mechanics. A monofractal refers to the case where $\{\alpha\}$
is a singleton. The singularity spectrum is called multifractal when $\{
\alpha\}$ spans an interval.

\subsection{WTMM-based JPFA of Fractal Correlation}

In this section, the WTMM approach is generalized and applied in the JPFA
of fractal correlation between multifractal singularity spectra. We present
the application of JPFA of two data sets. The extention to more data sets
is conceptually similar.

Consider $x_1(t),x_2(t)$ and their respective sets of singularity exponent
$\{\aa\},\{\ab\}$. Let the maxima lines of $|\T[x_k]|$ at scale $a$ be
denoted as ${\cal L}_k(a),k=1,2$. A natural extension of the existing WTMM
analysis is to consider a joint partition function of the form:
$$Z(a;q_1,q_2)=\sum_jC_{1,r(j)}^{q_1}C_{2,s(j)}^{q_2}
\eqno(5)$$
where $C_{1,r},C_{2,s}$ are the modulus maxima along the maxima lines $l_{
1,r}\in{\cal L}_1,l_{2,s}\in{\cal L}_2$.

To realize (5), the maxima lines in ${\cal L}_k,k=1,2$ must be paired up
properly (so the index $j$ can run). As in most correlation analyses, the
goal is to characterize the property related to observing both singularity
exponents $\aa$ and $\ab$. In terms of the WTMM analysis, such information
is contained in the modulus of the neighboring maxima lines. If the time
coordinate of $l_{k,j}(a)$ is denoted by $t_{k,j}(a)$, this means the
coefficients $C_{1,r}$, $C_{2,s}$ paired up in (5) can be determined by
$$|t_{1,r}-t_{2,s}|=\min_{r'}(|t_{1,r'}-t_{2,s}|)
=\min_{s'}(|t_{1,r}-t_{2,s'}|)
\eqno(6)$$

Once (5) and (6) are established, similar procedure developed by Bacry et
al. can be extended
to characterize the geometry associated with the
observation of $\aa$ and $\ab$. In particular, based on $C_{k,\lambda}\sim
a^{\alpha_k(\lambda)}$, $\lambda=r,s$ [15,17,18], (5) can be given by
$$Z(a;q_1,q_2)\sim\sum_j a^{q_1\alpha_1(r(j))+q_2\alpha_2(s(j))}
=\int\int d\aa d\ab {\cal P}(\aa,\ab)a^{q_1\aa+q_2\ab}a^{-f(\aa,\ab)}
\eqno(7)$$
where ${\cal P}(\aa,\ab)$ and $f(\aa,\ab)$ are the probability density
function and Hausdorff dimension of the support of $(\aa,\ab)$,
respectively. Applying the standard argument of steepest descent in small
$a$, one has
$$Z(a;q_1,q_2)\sim a^{\tau(q_1,q_2)}
\eqno(8)$$
where 
$$\tau(q_1,q_2)=\min_{\aa,\ab}(q_1\aa+q_2\ab-f(\aa,\ab)).
\eqno(9)$$
Hence, $\tau(q_1,q_2)$ and $f(\aa,\ab)$ are Legendre transform pair:
$$\aa=\partial\tau(q_1,q_2)/\partial q_1,
\ab=\partial\tau(q_1,q_2)/\partial q_2,$$
$$f(\aa,\ab)=\aa(q_1,q_2)q_1+\ab(q_1,q_2)q_2-\tau(q_1,q_2).
\eqno(10)$$
where
$$q_1=\partial f/\partial\aa, q_2=\partial f/\partial\ab.
\eqno(11)$$
Finally, from (9)~$\sim$~(11), the correlation coefficient between $\aa,\ab$
can be estimated using $\tau(q_1,q_2)$:
$$\rho={{\rm cov}(\aa,\ab)\over\sigma_\aa\sigma_\ab}=
-{{\partial^2\tau\over\partial q_1\partial q_2}\over
\sqrt{\left[{\partial^2\tau\over\partial q_1^2}
{\partial^2\tau\over\partial q_2^2}\right]}}\Biggr|_{q_1=q_2=0}
\eqno(12)$$
where cov denotes the covariance and $\sigma_\lambda$ denotes the standard
deviation of $\lambda$. This expression will be used in the next section to
compare with the numerical result.

In practice, the Legendre transform (10) relies on using $\tau$ estimated
from (8). However, there are known factors, such as lacunarity [19, 20],
that introduce oscillatory, scale dependent, prefactor. This results in
the poor estimate of $\tau(q_1,q_2)$. A remedy to this problem can be
motivated by an alternative approach equivalent to the canonical ensemble
in statistical mechanics [20]. Let
$$\nu(j,a;q_1,q_2)={C_{1,r(j)}^{q_1}C_{2,s(j)}^{q_2}\over Z(a;q_1,q_2)}.
\eqno(13)$$
Then, it can be shown (Appendix):
$$\A_1(a;q_1,q_2) = \sum_j\nu(j,a;q_1,q_2)\log(C_{1,r(j)})\sim a^{\aa(q_1,q_2)}
\eqno(14)$$
$$\A_2(a;q_1,q_2) = \sum_j\nu(j,a;q_1,q_2)\log(C_{2,s(j)})\sim a^{\ab(q_1,q_2)}
\eqno(15)$$
$$\F(a;q_1,q_2) = \sum_j\nu(j,a;q_1,q_2)\log(\nu(j,a;q_1,q_2))\sim
a^{f(\aa,\ab)}.
\eqno(16)$$

\subsection{JPFA and Relative Multifractal Analysis}

Relative multifractal analysis and similar ideas were developed to
characterize fractal correlation between fractal measures. The main idea
is to replace the use of Lebesgue measure in the traditional fractal
analysis [5,6,7,8]. Specifically, consider the partition functions of
multifractal measures $\pi$ and $\mu$
$$\sum_{A\in\cal H}\pi(A)^q\sim |A|^{\tau_\pi(q)}, \ \ \ \
\sum_{A'\in\cal H'}\mu(A')^q\sim |A'|^{\tau_\mu(q)}
\eqno(17)$$
where ${\cal H, H'}$ denote generic partitions of the support and $|\cdot|$
denote the Lebesgue measure of the set. To examine the extent to which the
singularity of $\pi$ correlate with $\mu$, the sets which scale as a power
law will now be characterized by using $\pi$. For example, the partition
function of $\pi$ is now written as
$$\sum\pi(A)^q\mu(A)^{-{\bf t}(q)}\sim O(|A|)
\eqno(18)$$
where the ``big $O$" describes the order relationship $O(|A|)\to{\rm const.
}$ as $|A|\to 0$. Define $\tau_{\pi/\mu}(q)=\sup\{{\bf t}(q)\}$ for which
(18) holds. The relative multifractal spectrum is obtained via the Legendre
transform of $\tau_{\pi/\mu}$. It characterizes the support of the singular
behaviour of the form $\pi\sim\mu^{\alpha_{\pi/\mu}(q)}$ where $\alpha_{\pi
/\mu}(q)=d\tau_{\pi/\mu}(q)/dq$. The relative multifractal analysis can
draw a much sharper distinction between $\pi$ and $\mu$. For example, $\tau
_{\pi/\mu}(q)$ is nonlinear when $\pi\ne\mu$ and $\tau_{\pi/\mu}(q)=q-1$
when $\pi=\mu$; see [6] for more details.

Comparison of (18) with (5) and (8) suggests $\tau_{\pi/\mu}$ can be
obtained as the level set of $\tau(q_1,q_2)=0$ where
$$q_2=-\tau_{\pi/\mu}(q_1).
\eqno(19)$$
To assure finite generalized dimension, $\tau_{\pi/\mu}(1)=0$ and, thus,
the level set $\tau(q_1,q_2)=0$ must pass through $(q_1,q_2)=(1,0)$. This
property can also be directly seen for the multinomial processes (next
section).

Similarly, by switching the role of $q_1,q_2$, the singular behaviour of
$\mu$ ``viewed" by $\pi$ can be described. With the same argument, this is
characterized by $\tau_{\mu/\pi}(q_2)$ defined from the same level set
$\tau(q_1,q_2)=0$ where $q_1=-\tau_{\mu/\pi}(q_2)$. It may be useful to
point out that $\tau_{\pi/\mu},\tau_{\mu/\pi}$ on the $q_1\times q_2$ plane
are nothing but mirror images of the contour of $\tau(q_1,q_2)=0$ about
$q_2=0$ and $q_1=0$ axes, respectively. In general, $\tau_{\pi/\mu}(q)\ne
\tau_{\mu/\pi}(q)$, although they are derived from the same level set.

\section{Numerical Experiments}

To test if the WTMM-based JPFA can reliably characterize fractal correlation,
numerical experiments are conducted on the coupled random binomial cascades
studied by Meneveau et al. [1].

The first cascade, denoted as $\pi$, is generated by weights $\pa,\pb$
(referred to as $\pi$-cascade). Let $I_{r_1,\cdots,r_J}$ denote an interval
segment generated in the $J$th iteration where $r_i\in\{0,1\}$ and $\sum
r_i2^{-i}$ is the based-2 coarse-grained representation of any $x\in I_{
r_1,\cdots,r_J}$. By the multiplicative rule, $\pi(I_{r_1,\cdots,r_J})=
\prod_{j=1}^Jp_{r_j}$. The second cascade, denoted as $\mu$, is generated
by weights $\ma,\mb$ (referred to as the $\mu$-cascade). With the same
addressing scheme, one has $\mu(I_{s_1,\cdots,s_J})=\prod_{j=1}^Jm_{s_j}$
where $s_i\in\{0,1\}$.

In the numerical experiment, a parameter $g$ and a uniform random variable
$\gamma$ in [0,1] are used to control the degree of coupling or correlation
between the cascades. Let $I_L$, $I_R$ be the new segments created from
their parent segment of the previous generation. If $\gamma<g$, the weights
assigned to $I_L,I_R$ of the $\mu$-cascade will depend on exactly how the
weights of the $\pi$-cascade are assigned. The rule for this dependence is
that $\pa$ and $\ma$ ($\pb$ and $\mb$) are always assigned at the same time.
For example, if $\pa$ is assigned to $I_L$ ($I_R$) of the $\pi$-cascade,
$\ma$ will be assigned to $I_L$ ($I_R$) of the $\mu$-cascade and similarly
for $\pb$ and $\mb$. If $\gamma\ge g$, the weight assignment for the
cascades will be completely independent from each other. This way, the
fractal generating mechanisms of the cascades are completely dependent of
each other when $g=1^+$ and independent of each other when $g=0^-$.

The coarse-grained joint partition function for the coupled cascades can
be defined based on (5):
$$Z_J(a;q_1,q_2)=\sum\pi(I_{r_1,\cdots,r_J})^{q_1}
\mu(I_{r_1,\cdots,r_J})^{q_2}.$$
From the combination of $\gamma$ completely dependent and $(1-\gamma)$
independent proportions, $Z_J$ is derived explicitly as
$$Z_J(a;q_1,q_2)\sim (2Y)^J
\eqno(20)$$
where
$$Y=\gamma\left(\pa^{q_1}\ma^{q_2}+\pb^{q_1}\mb^{q_2}\over 2\right)
+(1-\gamma)\left(\pa^{q_1}\ma^{q_2}+\pb^{q_1}\ma^{q_2}+
\pa^{q_1}\mb^{q_2}+\pb^{q_1}\mb^{q_2}\over 4\right).
\eqno(21)$$
As $J\to\infty$, $Z_J\to Z$ and the analytical $\tau(q_1,q_2)$ can be
obtained from (21) as
$$\tau(q_1,q_2)=-\log_2(2Y)
\eqno(22)$$
By (10), the analytical $\aa,\ab$ and $f(\aa,\ab)$ can be found [21].
These results will be compared to the numerical ones below.

In the numerical experiments, $\pa=0.2,\pb=0.8$ and $\ma=0.4,\mb=0.6$ are
used to generate the $\pi$- and $\mu$-cascades for $g=1,0.8,0.3,0$. For
each $g$ value, 30 pairs of $\pi$, $\mu$ cascades, each of 16,384 points
are generated. The first derivative of the Gaussian wavelet has been used
as the analyzing wavelet in this work. Using higher order derivative of
the Gaussian wavelet does not create  qualitatively different result. In
practice, the modulus maxima and the maxima lines of the individual cascade
are first obtained. The modulus maxima from the nearest maxima lines are
then paired up according to (6) and used in (5) to define the joint
partition function. The numerical $\aa,\ab$ and $f(\aa,\ab)$ are finally
estimated following (13)~$\sim$~(16) and $\tau(q_1,q_2)$ is determined
following (9).

Typical maxima lines of the coupled cascades are shown in Fig.~2. It is
observed that the maxima lines are ``aligned" when the fractal generation
is completely dependent at $g=1$ and begin to ``mis-align" for $g<1$. The
power law scaling of $Z(a;q_1,q_2)$ are found in all cases (Fig.~3). In
Fig.~4, contours of the level set of $f(\aa,\ab)$ are shown on the $\aa
\times\ab$ plane. It is evident that the geometry of the contour lines vary
systematically with the $g$ value. When the fractal generating mechanisms
are completely dependent of each other ($g=1$), $f(\aa,\ab)$ describes a
one-dimensional curve supported by the functional relationship $\aa(\ab)$.
This is expected as any spiking pattern in one cascade automatically
implies the same for the other. As a result, the maxima lines will converge
at the same location in the time-scale plane. This establishes the
one-to-one relationship of observing the exponents $\aa$ and $\ab$. In
general ($g<1$), $f(\aa,\ab)$ describes a two-dimensional surface, which
gives rise to oval-shape contours (Figs.~4b, 4c, 4d). This means that the
observation of $\aa$ can take place simultaneously for a range of $\ab$. As
a result, the contour ``opens up" and becomes the largest when the fractal
generations are completely independent from each other ($g=0$). For $g=0$,
there is a perfect alignment of the axes of the contour and the $\aa=0$ and
$\ab=0$ axes (Fig.~4d). Superimposed on these figures are the analytical
$f(\aa,\ab)$ derived by the Legendre transform of (22). It is seen that the
WTMM-based JPFA agrees well with the theory. To further the check of the
WTMM-based approach, the correlation of $\aa, \ab$ is estimated using the
numerical $\tau(q_1,q_2)$. This is to compare with the analytical $\rho=g$
obtained by substituing (22) into (13). The result is summarized in Table 1.
Again, good agreement is found.

\bigskip

\begin{center}
\begin{tabular}{l|l|l|l|l}
\hline
$g$&0.0&0.3&0.8&1.0\\
\hline
$\rho$&0.016&0.280&0.775&0.838\\
\hline
\end{tabular}
\end{center}

\begin{center}
TABLE 1 Numerical $\rho$ value for $g=0,0.3,0.8,1.0$ (Note, in theory, $\rho=g$).
\end{center}

\bigskip

To test the robustness of the algorithm, different analyzing wavelets are
also used to study the coupled cascades. While deviations are expected to
result from the choice of the analyzing wavelets, no qualitatively
different result is found. Fig.~5 demonstrates the coupled cascades of $g
=0.8$. It is evident that the contour lines of $f(\aa,\ab)$ estimated
from different analyizing wavelets are all falled onto the theoretical
contour lines.

Finally, from the numerical data, $\tau_{\pi/\mu}(q_1)$ is estimated from
the level set $\tau(q_1,q_2)=0$. We then conduct the relative multifractal
analysis using the deterministic algorithm proposed by Riedi and Scheuring
[6]. In Fig.~6, the $\tau_{\pi/\mu}(q_1)$ estimated by these two different
approaches are shown to match well. The agreement confirms that the relative
multifractal spectrum is contained in the level sets of $\tau(q_1,q_2)$.
Similar match is also found for $\tau_{\mu/\pi}$ (not shown).

\section{Concluding Remarks}

In applications, the ability to characterize fractal correlation in the
data fluctuation could provide insights into the underlying complex dynamics.

In this work, a WTMM-based technique is introduced for the first time to
estimate the fractal correlation in the framework of joint partition
function analysis proposed by Meneveau et al. [1]. As WTMM has been proven
an effective tool to extract the singularity spectrum of certain important
class of fractal functions [15, 16], it is shown that the extension
developed in this work also capture accurately the fractal correlation of
data fluctuation. We also found another leading idea developed for the
fractal correlation analysis, relative multifractal spectrum, can be
considered as a special case of JPFA at a particular parameter setting.

\bigskip
\noindent{\bf Acknowledgment}
\bigskip
This research is supported by Natural Science and Engineering Research
Council of Canada.

\bigskip
\begin{center}
\noindent{\bf APPENDIX}
\end{center}
\bigskip

First, consider (14). By (8) and (10), one has
$${\partial Z(a;q_1,q_2)/\partial q_1}\sim a^{\tau(q_1,q_2)}\log(a)
{\partial\tau(q_1,q_2)/\partial q_1}=a^{\tau(q_1,q_2)}\log(a)\alpha_1.
\eqno(A.1)$$
Note the prefactor $\log(a)\alpha_1$ in $(A.1)$ that varies logarthmically
with $a$. From $(A.1)$, one has
$$\sum_j\nu\log(|C_{1,j}|)={\partial Z/\partial q_1\over Z}.
\eqno(A.2)$$
Equation (14) follows after substituting $(A.1)$ into $(A.2)$. Note also
the prefactor in (8) is canceled out in $(A.2)$. The derivation for (15)
is the same.

Based on $(A.2)$, one has 
$$\sum\nu\log(\nu)=q_1{\partial Z/\partial q_1\over Z}
                  +q_2{\partial Z/\partial q_2\over Z}-\log(Z).
\eqno(A.3)$$
Again, using $(A.1)$, $(16)$ results after substituting (10) into the
above.

\bigskip
\noindent{\bf Reference}
\bigskip

\noindent [1] C. Meneveau et al., {\it Phys. Rev. A.} {\bf 41}, 894 (1990).

\noindent [2] J.O. Fortrat et al., {\it Auton. Neurosci.} {\bf 86}, 192
(2001); also general review in Task Force of the ESC and NASPE, {\it Euro.
Heart J.} {\bf 17}, 354 (1996).

\noindent [3] J. L${\acute e}$vy-L${\acute e}$hel and R. Riedi, Fractal in
Engineering, Eds. J. L${\acute e}$vy-L${\acute e}$hel, E. Lutton, C. Tricot,
{\it Springer Verlag}, London, 185 (1997).

\noindent [4] Y. Pesin, Dimension theory in dynamical systems: contemporary
views and applications, Chicago Lectures in Mathematics, {\it Chicago
Univ. Press} (1997).

\noindent [5] J. L${\acute e}$vy-L${\acute e}$hel and R. Vojak, {\it Adv. Appl.
Math.} {\bf 20}, 1 (1998).

\noindent [6] R.H. Riedi and I. Scheuring, {\it Fractals} {\bf 5}, 153 (1997).

\noindent [7] G. Brown et al., {\it J. Stat. Phys} {\bf 66}, 775 (1992).

\noindent [8] J. Cole, {\it Chaos, Solitons \& Fractals} {\bf 11},
2233 (2000).

\noindent [9] H.G.E. Hentschel and I. Procaccia, {\it Physica D} {\bf 8},
435 (1983).

\noindent [10] M.H. Halsey et al., {\it Phys. Rev. A} {\bf 33}, 1141 (1986).

\noindent [11] D. Schertzer and S. Lovejoy, Space/Time Variability and
Interdependencies in Hydrological Processes, Ed. R.A. Feddes, {\it
Cambridge Univ. Press}, 153 (1995).

\noindent [12] A.N. Kravchenko et al., {\it Agron. J.} {\bf 92}, 1279 (2000).

\noindent [13] T.B. Zeleke et al., {\it Soil Sci. Soc. Am. J.} {\bf 69}, 1691
(2005).

\noindent [14] W-X. Zhou and D. Sornette, arXiv:cond-mat/0408600 (2004).

\noindent [15] E. Bacry et al., {\it J. Stat. Phys} {\bf 70}, 635 (1993).

\noindent [16] S. Jaffard, {\it SIAM J. Math. Anal.} {\bf 28}, 944 (1997).

\noindent [17] S. Mallat and W.L. Hwang, {\it IEEE Trans. Info. Theory}
{\bf 38}, 617 (1992).

\noindent [18] J.F. Muzy et al. {\it Phys. Rev. E} {\bf 47}, 875 (1993).

\noindent [19] C. Meneveau and K.R. Sreenivasan, {\it Phys. Lett. A} {\bf
137}, 103 (1989).

\noindent [20] A. Chhabra and R.V. Jensen, {\it Phys. Rev. Lett.} {\bf
62}, 1327 (1989).

\noindent [21] A. Sharif, {\it Master Thesis}, Department of Mech. \&
Ind. Eng., Ryerson University (2006).

\vfill\eject
\noindent{\bf Figure Captions}
\bigskip

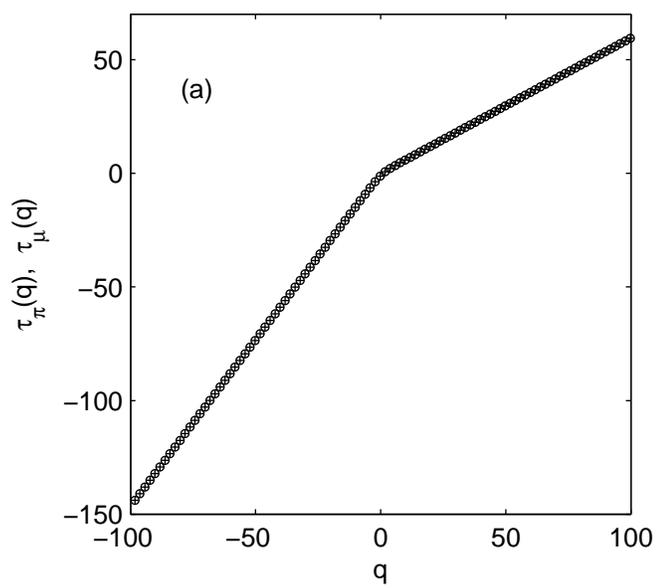
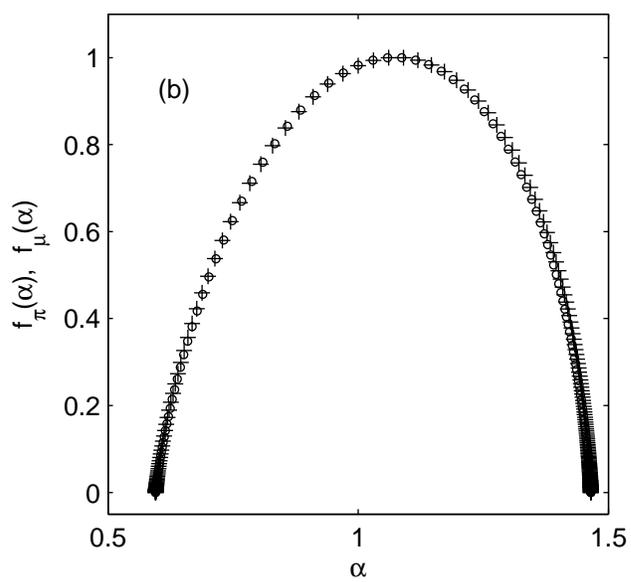
\noindent Fig.~1. Multifractal analyses of 3-adic $\pi$-cascade of weights
0.2, 0.28, 0.52 (symbol "o") and 5-adic $\mu$-cascade of weights 0.09463,
0.1200, 0.1800, 0.2217, 0.3837 (symbol "+"). (a) $\tau_\pi(q)$ and $\tau_
\mu(q)$; (b) $f_\pi(\alpha)$ and $f_\mu(\alpha)$.
\bigskip

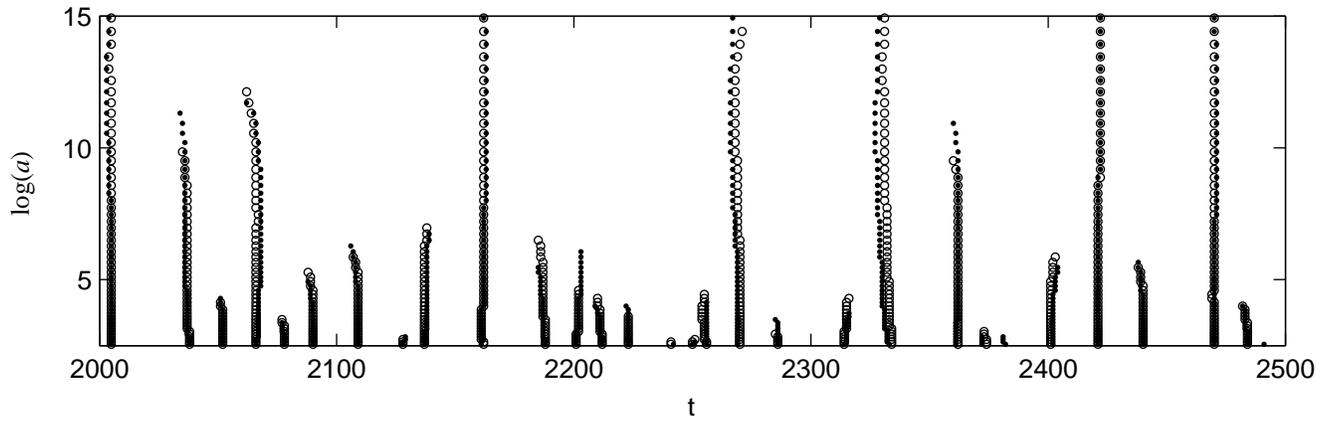
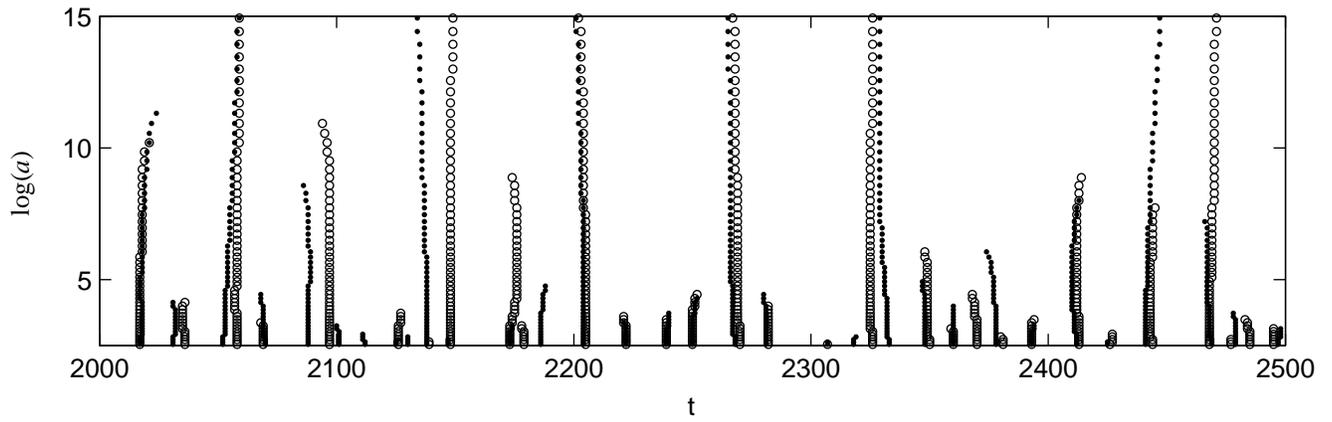
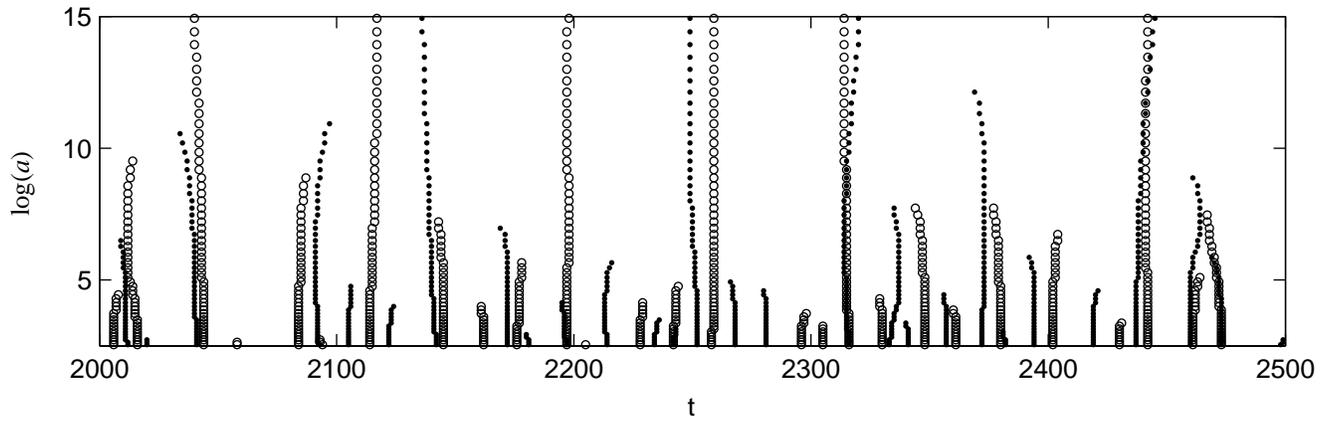
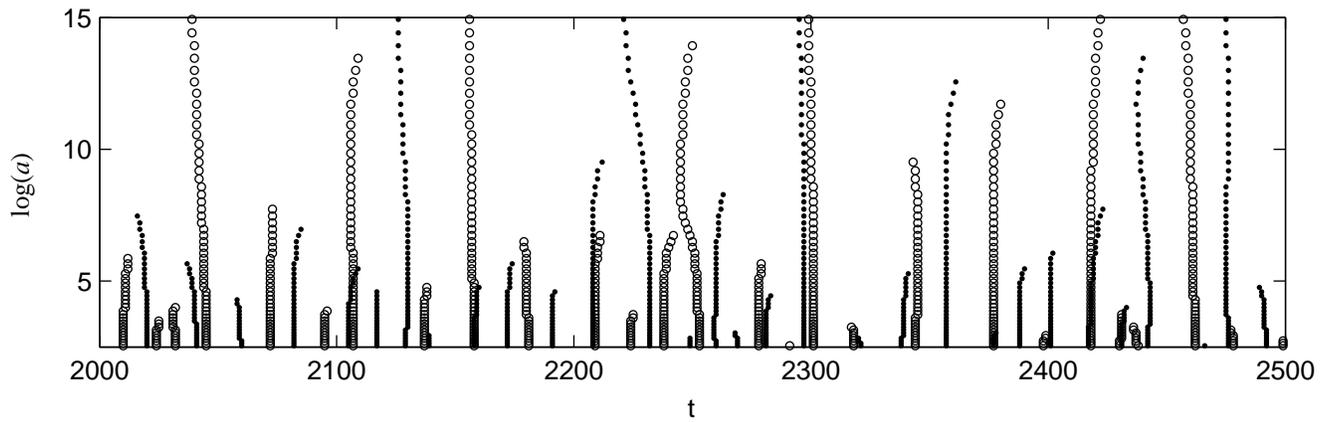
\noindent Fig.~2. Typical maxima lines in the time-scale plane from one
of the 30 sets of $\pi$- ("o") and $\mu$-cascades ("+") with coupling
parameter $g=1.0, 0.8, 0.3, 0.0$ (top to bottom). Notice the perfect
alignment of maxima lines for the completely dependent cascades ($g=1$).
First derivative of the Gaussian wavelet is used in the numerical
calculation.
\bigskip

\noindent Fig.~3 $\log({\bf A}_1)$, $\log({\bf A}_1)$ and $\log({\bf F})$
vs. $\log(a)$ plots of a typical case of the coupled cascades with $g=
0.8$. The straight lines describes the power laws at $(q_1,q_2)$ = (3,-2),
(4,0), (0,0), (-1,3) (top to bottom). Regression lines are shown as solid
lines. The slope of the regression lines are estimated as $\aa$, $\ab$ and
$f(\aa,\ab)$ based on (14)--(16).
\bigskip

\noindent Fig.~4 Averaged contour lines of numerical $f(\aa,\ab)=C$ for
$C=0.6$ (outer contour), 0.75, 0.9 (inner contour). Theoretical contour
lines are shown as solid lines. The averaging over 30 pairs of $\pi$ and
$\mu$ cascades is shown as ``o" ($C=0.6$), ``$\bullet$" ($C=0.75$) and
``$\triangle$" ($C=0.9$). Error bars of one standard deviation from
selected data points are shown. They are obtained from the ensemble of
contour line points in a uniform grid on the $q\times  p$. The
corresponding $g$ values are (a) 1, (b) 0.8, (c) 0.3, (d) 0.
\bigskip

\noindent Fig.~5 Averaged contour lines of numerical $f(\aa,\ab)=C$ for
$g=0.8$ and $C=0.6$ (outer contour), 0.75, 0.9 (inner contour). The
analyzing wavelets are ... (``o"), ... (``+"), ... (``$\triangle$").
Theoretical contour lines are shown as solid lines. Error bars of one
standard deviation from selected data points are shown. Compare with
Fig.~4b.

\noindent Fig.~6 $\tau_{\pi/\mu}(q)$ estimated by WTMM-based JPFA ('o')
method (from the contour line of $\tau(q,p)=0$) and the deterministic
algorithm porposed (solid line); see text. The solid lines shown are
based on the ensemble average with $\pm$ 3 standard deviation boundaries
plotted as long-dashed lines. The $g$ values are (a) 1, (b) 0.8, (c) 0.3,
(d) 0.

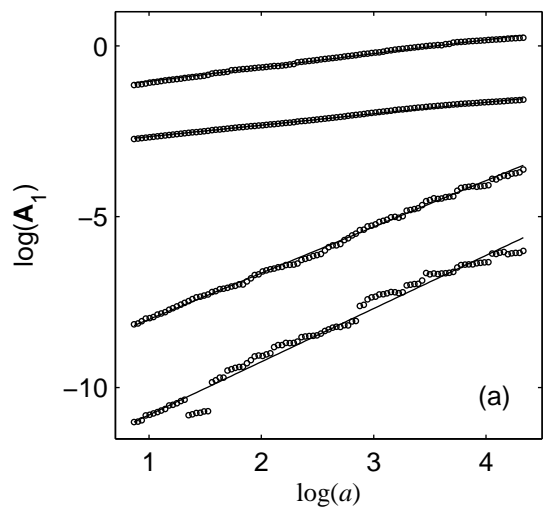
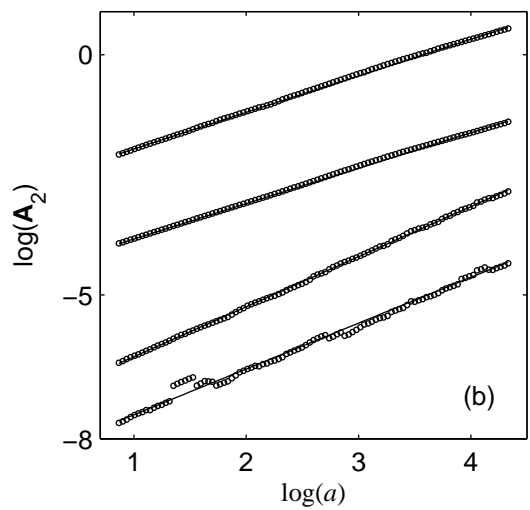
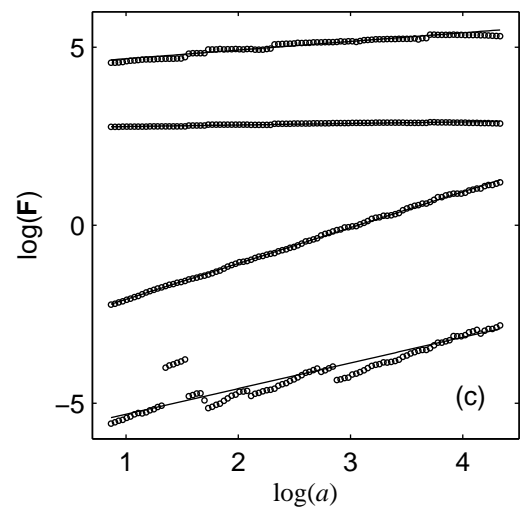
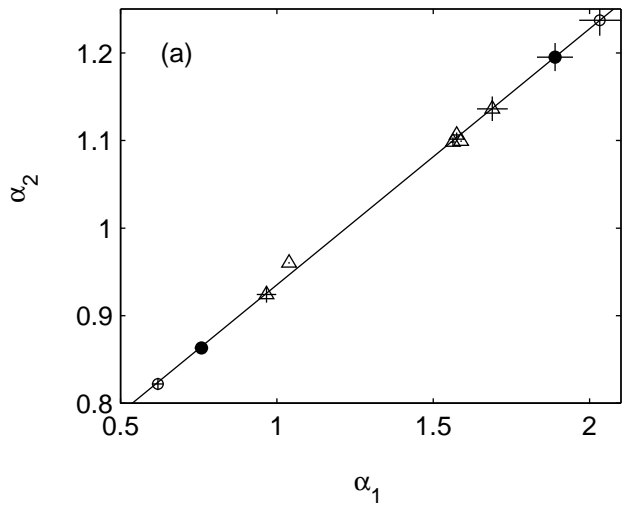
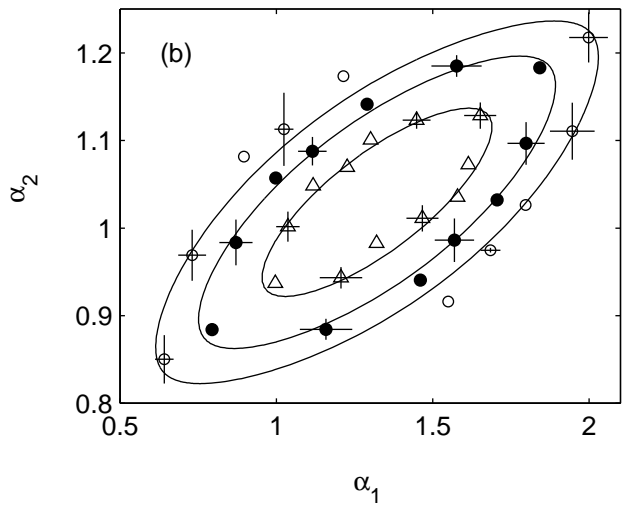
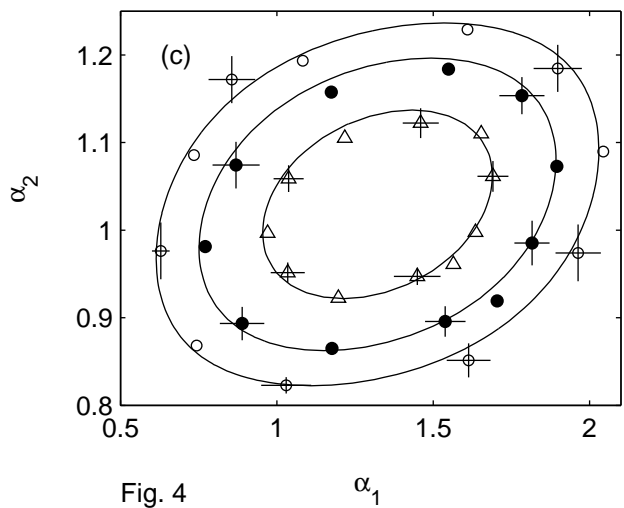
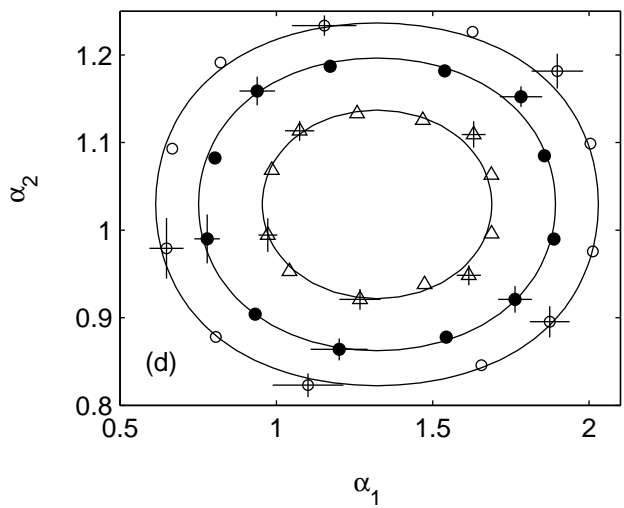
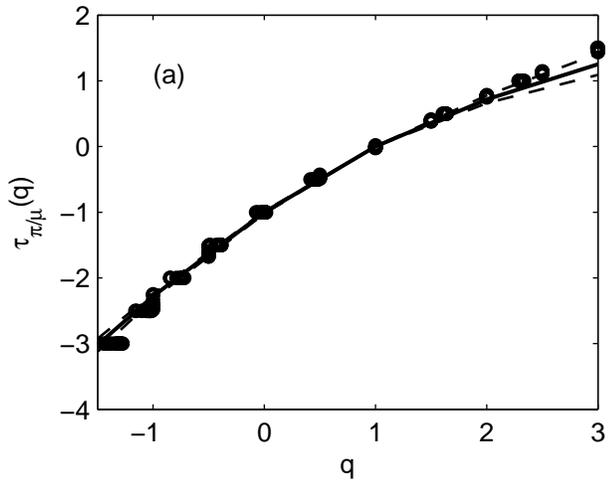
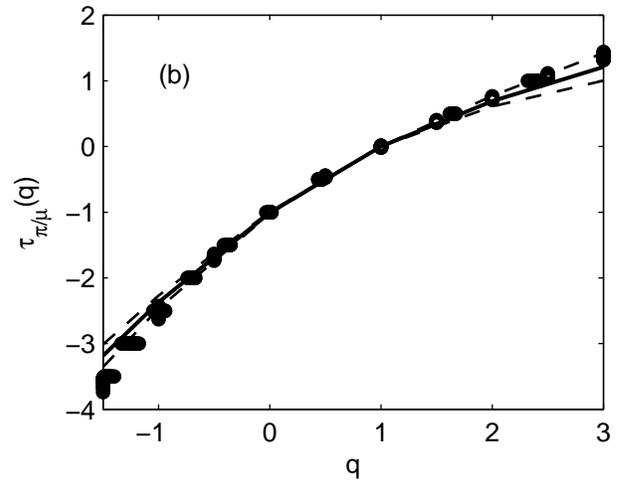
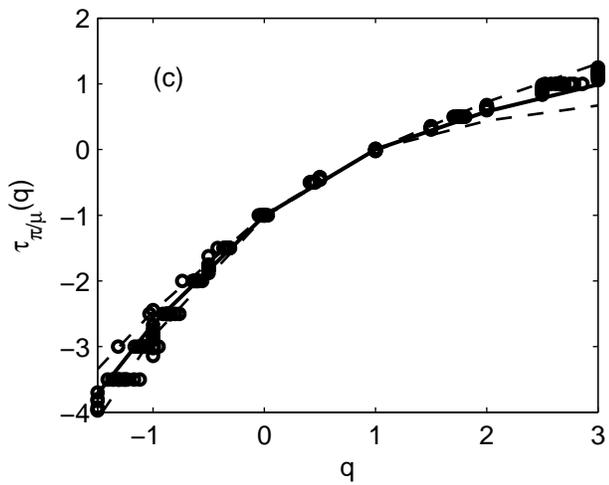
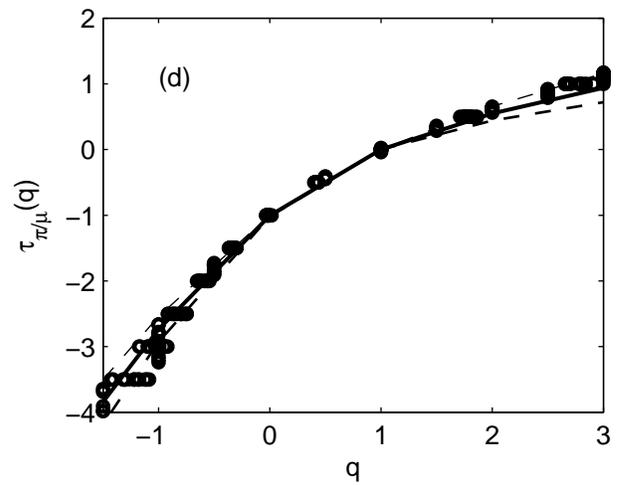
\end{document}